\documentclass[authoryear,12pt]{article}
\newcommand{\bd}[1]{\ensuremath{\mbox{\boldmath $#1$}}}
\usepackage{natbib}
\usepackage{graphicx}
\usepackage{geometry}
\usepackage{amsmath}
\usepackage{setspace}
\providecommand{\keywords}[1]{\textbf{\textit{Keywords:}} #1}
\begin{document}
\begin{spacing}{2}
\title{Growth Mixture Modeling with Measurement Selection}

%
%

%
%

\author{Abby Flynt\\
Assistant Professor\\
Department of Mathematics\\
Bucknell University\\
Lewisburg, PA 17837, USA\\
abby.flynt@bucknell.edu\\
\and
Nema Dean\\
Lecturer\\
School of Mathematics and Statistics\\
University of Glasgow\\
Glasgow G12 8QQ, UK\\
 nema.dean@glasgow.ac.uk\\
 \date{}
}

\normalsize
\date{}
\maketitle

\abstract{Growth mixture models are an important tool for detecting group structure in repeated measures data. Unlike traditional clustering methods, they explicitly model the repeat measurements on observations, and the statistical framework they are based on allows for model selection methods to be used to select the number of clusters. However, the basic growth mixture model makes the assumption that all of the measurements in the data have grouping information/separate the clusters. In other clustering contexts, it has been shown that including non-clustering variables in clustering procedures can lead to poor estimation of the group structure both in terms of the number of clusters and cluster membership/parameters. In this paper, we present an extension of the growth mixture model that allows for incorporation of stepwise variable selection based on the work done by \citet{maugis09} and \citet{raftery06}. Results presented on a simulation study suggest that the method performs well in correctly selecting the clustering variables and improves on recovery of the cluster structure compared with the basic growth mixture model. The paper also presents an application of the model to a clinical study dataset and concludes with a discussion and suggestions for directions of future work in this area.}

\section{Introduction}
Cluster analysis is the search for group structure in multivariate data where little or no \emph{a priori} information about groups is available \citep{everitt11}. There are a wide variety of different types of cluster analysis approaches available. These can be broadly categorized into three classes: algorithmic (which includes k-means \citep{macqueen67} and hierarchical clustering \citep{ward63}), non-parametric (mode-hunting, cluster tree approaches, \citep{wishart69}, Section 11 in \citep{hartigan75} and \citep{hartigan81}) and parametric (finite mixture model clustering \citep{titterington85}). Finite mixture model clustering (also commonly known as model-based clustering) is becoming more popular in many application areas due to the wide availability of software, ease of interpretation of output and limited number of subjective decisions necessary for its application. 

This paper will focus on growth mixture modeling which is a special case of finite mixture model clustering. 
Growth mixture modeling is a framework that allows for cluster detection in situations with repeated measurements. It was first introduced by \citet{muthen99} and a good review can be found in \citet{ram2009}. 
The growth mixture model (GMM) framework allows for modeling of the repeated measurements either directly or as a regression model of outcome measurements on explanatory measurements.

\begin{figure}[htbp]
\begin{center}
\includegraphics[scale=0.61]{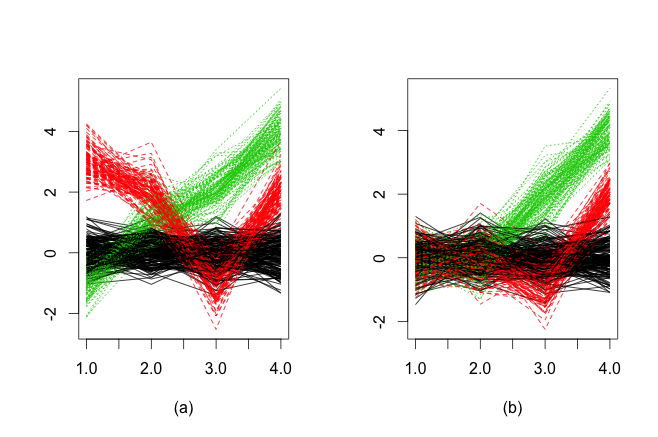}
\end{center}
\caption{(a) 3 groups, where all time-points along x-axis are important for separating groups; (b) 3 groups, where only 3rd and 4th time-points are important for separating groups. Different groups have different line types and colors.}\label{fig:intro}
\end{figure}

One assumption made by the GMM is that all of the repeated measurements are important to the group structure. Figure \ref{fig:intro} (a) shows an example where this is the case, where each repeated measure/time point has a different mean for each group. In cases where the GMM at each time point is a separate regression, this would mean each component in the GMM had a different intercept and/or slope for each group at each repeated measure. However, if the levels or slopes/intercepts at some measurements do not vary across groups, we may have a situation similar to Figure \ref{fig:intro} (b). In the latter example, only the last two repeated measurements are important for separating the three groups. So the assumption of all measurements having clustering information may not be true and if noise variables are included, it has been shown in other contexts \citep{raftery06, rusakov05} that this can be detrimental to the performance of the clustering method. It is also the case that from a substantive point of view, knowing which measurements/time points differentiate between groups may be of interest in itself. For example, Figure \ref{fig:intro} (b) could be a case where a medication does have differing impacts on groups of patients but this difference effect is delayed and not visible in the first two measurements/time points. This paper applies the variable selection method proposed in \citet{raftery06} and \citet{maugis09} to the GMM to simultaneously allow for selection of the repeated measurements that drive the clustering with application of the clustering itself. Simultaneous or ``wrapper'' selection of clustering variables along with cluster model estimation is generally preferable to either ``filter'' approaches that select variables prior to clustering or post hoc approaches that select variables after the cluster model has been estimated. There is something of a chicken-and-egg problem with variable selection in clustering, as the variables included often drive the clustering found, and the clustering estimated defines the variables of interest. As such, a simultaneous approach is to be preferred to try to tackle both problems at the same time.

We begin with the methods in Section \ref{sec:methods}, where in Section \ref{sec:gmm}, we introduce the general GMM and discuss its properties and estimation. We then summarize the variable selection framework of \citet{raftery06} in Section \ref{sec:varsel}. The specific variable selection for both a basic GMM (without cluster-specific measurement regression) and the regression GMM is explained in \ref{sec:varselgmm}. These are applied in Section \ref{sec:results} to a variety of different settings in a simulation study, with the results presented in Section \ref{sec:simstudy}, followed by results on the Pittsburgh 600 dataset in Section \ref{sec:data}. The paper wraps up with a summary of the main results, some caveats and future directions for further research on this topic in Section \ref{sec:discuss}.

\section{Methods}\label{sec:methods}
\subsection{Finite Mixture Model Clustering}\label{sec:fmm}
One of the first major analyses using a finite mixture model was introduced by \citet{pearson94}. Finite mixture model clustering assumes that instead of a single homogenous population, data may come from a heterogenous population made up of homogenous sub-populations. Rather than directly model the overall population with a single density, each sub-population is modeled with its own density. The overall population density can then be expressed as a weighted sum of the component/sub-population densities, with the weights set to the proportions each sub-population makes up of the overall population.

So, if we have an individual (multivariate) observation $\bd{y}$, the mixture density with $K$ components/sub-populations is given by:
\begin{equation} f(\bd{y}|\bd{\theta}) = \sum_{k=1}^K\pi_kf_k(\bd{y}|\bd{\theta}_k), \label{eqn:mixmod}\end{equation}
where
\[ \pi_k \geq 0, \,\,\, \sum_{k=1}^K\pi_k = 1.\]
Here, the $\pi_k$'s are the mixture proportions and the $\bd{\theta}_k$ are the sets of parameters for each component density. If $\bd{y}$ is continuous, then $f_k$ is often taken to be Gaussian and then $\bd{\theta}_k = (\bd{\mu}_k, \bd{\Sigma}_k)$, a set of component specific mean vectors, $\bd{\mu}_k$, and component specific covariance matrices, $\bd{\Sigma}_k$. Clustering using finite mixture models with Gaussian component densities is commonly known as model-based clustering (see \citet{fraley98} for details).

If we have a single variable $y$ that is related to another variable $x$ or a vector of variables $\bd{x}=(x_1, \ldots, x_p)$, then we can cluster the relationship between $y$ and $\bd{x}$ via a finite mixture of regression models as given by:
\begin{equation} f(y|\bd{x},\bd{\theta}) = \sum_{k=1}^K\pi_k f_k(y|\bd{x},\bd{\theta}_k)\label{eqn:mixreg}\end{equation}
Here, instead of having the marginal distributions of $y$ in the component densities (as with $f_k(\textbf{y}|\bd{\theta}_k)$ in \eqref{eqn:mixmod}), we have conditional component densities of $y$ given $\mathbf{x}$, $f_k(y|\mathbf{x},\bd{\theta}_k)$,  which represent component level regressions giving an overall finite mixture of regressions.

For continuous $y$, $f_k$ is usually assumed to be Gaussian distributed, with $\bd{\theta}_k = (\bd{\beta}_k,\sigma_k^2)$, a set of component specific regression parameter vectors $\bd{\beta}_k = (\beta_{0k},\beta_{1k},\ldots,\beta_{pk})$ and component specific variance $\sigma_k^2$. The $\pi_k$'s remains as before.
Regression relationships which are more complex than linear can be used and for a non-continuous $y$, more complex models like generalized linear models (see \cite{grun08} for details) or splines (see \cite{james03}) can be used. 

Estimation of the finite mixture model can be performed via either frequentist or Bayesian inferential methods. In the frequentist setting, the EM algorithm \citep{dempster77} or variants thereof \citep{gupta11, mclachlan08} is commonly used for estimation, where the observed data is augmented by missing data, in this case an identifier of which mixture component each observation is generated from. \citet{fraley02} gives details of model-based clustering estimation via EM.

\subsubsection{Choice of number of components}\label{sec:no comps}
Choosing the number of mixture components that best fits the data can be posed as a model choice question for finite mixture model clustering methods, in contrast to traditional algorithmic methods. Each number of components defines a different model so we can score the fit of each model, then choose the model (and associated number of components) that scores best. One of the most commonly used scoring mechanisms is the Bayesian Information Criterion, BIC \citep{schwarz78}. For a particular model $M$ with number of independent parameters $\nu$ and number of data points, $n$, the BIC can be defined as
\begin{equation} BIC(M) = -\log(\mbox{maximized likelihood of model }M) + \nu\log(n) \label{eqn:bic}\end{equation}
This essentially looks at how well the model fits the data, via the maximized log likelihood and penalizes the model complexity by the number of parameters required, weighted by the log of the number of observations. The best model will be the one with the \emph{smallest} BIC score.

Papers such as \cite{biernacki97} and \cite{biernacki99} have discussed using classification likelihood based criteria for model selection in the context of mixture modeling, but by far the most common measure used is the BIC (see \cite{fraley98} and \cite{fraley02} for further details), which has consistency results for selecting the correct order of the mixture model for Gaussian components under certain conditions (\cite{keribin00}). Thus, this is the criterion for component selection we look to use for the remainder of this paper. The approach could also be easily adapted to use a different method for choosing the orders of the mixtures.

\subsubsection{Assignment of observations to components}\label{sec:compassign}
For each observation $y$, instead of a hard assignment to a particular cluster, mixture model clustering can produce a vector of posterior class membership probabilities via Bayes' rule, using the fitted model parameters:
\begin{equation} \hat{p}_{sm} = P(\mbox{component } = m|y_s, \bd{x}_s, \bd{\hat{\theta}}) = \frac{\hat{\pi}_m f_m(y_s|\bd{x}_s,\bd{\hat{\theta}}_m)}{ \sum_{k=1}^K\hat{\pi}_k f_k(y_s|\bd{x}_s,\bd{\hat{\theta}}_k)} \label{eqn:classprob}\end{equation}

These class membership probabilities are one of the advantages of mixture model clustering. They can give a measure of uncertainty in the assignment of observations to components, which is not available from a hard clustering. They can also give an indication of the degree of overlap between components. 

If a hard classification is required, the \textit{maximum a posteriori} (MAP) mapping can be used. This assigns an object to the mixture component that has the highest value in the class membership probabilities. The MAP classification is simply:
\[ \mbox{Component for subject } s = \arg\max_{m}\hat{p}_{sm} \]

\subsubsection{Connecting a fitted mixture model to clustering}\label{sec:mixclust}
Once the best mixture model for the data has been selected, the most common approach to assigning clusters is that each fitted density component represents a cluster. This paper therefore assumes each GMM component found represents a cluster, however alternative approaches are discussed in Section \ref{sec:discuss}.

\subsection{Growth Mixture Model}\label{sec:gmm}
The growth mixture model is a mixture model applicable to data with multiple measurements, e.g. individuals recorded at multiple points over time. The assumption of conditional independence is made between different (sets of) repeated measurements, conditioned on component membership. This reduces the multivariate component density $f_k$ to a product of univariate component densities. 
If we have $S$ subjects/individuals, each with $N_s$ repeated measurements, the $k^{th}$ component density decomposes to the following product for multivariate observation $\bd{y}_s$: 
\begin{equation} f_k(\bd{y}_s|\bd{\theta}) = \prod_{n=1}^{N_s}f_{nk}(y_{sn}|\bd{\theta}_{nk}) ,\label{eqn:gmm}\end{equation}
or for the case of repeated measurements on an outcome variable $y$ and set of covariates $\bd{x}$ we have:
\begin{equation} f_k(\bd{y}_s|\bd{x}_s,\bd{\theta}) = \prod_{n=1}^{N_s}f_{nk}(y_{sn}|\bd{x}_{sn},\bd{\theta}_{nk}) ,\label{eqn:gmmreg}\end{equation}
where $n$ indexes the repeated measurements in each subject. 
This represents an assumption of independence between repeated measurements, conditional on the component membership. Latent class analysis makes a similar assumption of (component) conditional independence between categorical variables. Note that equation \eqref{eqn:gmm}, without covariates in the Gaussian setting, is a special case of the model-based clustering model with covariance parameterization set to be a diagonal matrix  within each component (where the diagonal elements are not required to be equal within or across components). 
This is the  ``VVI'' model in {\verb+mclust+} (model-based clustering package \citep{mclust,fraley02} in the {\verb+R+} software language \citep{Rlanguage}) parlance (where the volume and shape of clusters are allowed to ``V''ary across clusters but the orientation of the cluster ellipses is ``I''n parallel with variable axes).

We can use the EM algorithm estimation along with the standard methods as discussed in Sections \ref{sec:no comps}, \ref{sec:compassign} and \ref{sec:mixclust} to produce a clustering model in this framework.

\subsubsection{Mixed mode and missing data}\label{sec:missdata}
The conditional assumption means that the outcome variables can be mixed mode data, i.e.  of different types (continuous or categorical, binary versus count, etc.), as conditional independence means that maximization within the M step of the EM algorithm takes place over each repeated measurement separately. 
Thus, multiple outcome variables of different types (continuous/categorical) measured repeatedly can be jointly rather than separately modeled. Or outcomes that change from one type of variable to another over the time course, e.g. due to discretization, such as a blood pressure measure that is changed into high blood pressure status (yes/no), can be modeled in the same framework. The E step of the EM, essentially given by equation \eqref{eqn:classprob}, can easily be adapted to this situation as well, since it involves evaluation of each density separately before the product is taken. 

Similarly, if there is data missing for some of the repeated measurements for some subjects, the data from these subjects can still be used in the estimation of model parameters, and component memberships can still be calculated on these subjects. The maximization for a particular repeated measurement can take place over only the subjects for which there is no missing data on that measurement as a result of the conditional independence assumption. Similarly, equation \eqref{eqn:classprob} can be updated in the following way:
\[  \hat{p}_{sm} = P(\mbox{component } = m|y_s, \bd{x}_s, \bd{\hat{\theta}}) = \frac{\hat{\pi}_m f_m(y_s|\bd{x}_s,\bd{\hat{\theta}}_m)}{ \sum_{k=1}^K\hat{\pi}_k f_k(y_s|\bd{x}_s,\bd{\hat{\theta}}_k)}, \]
where 
\[ f_k(y_s|\bd{x}_s,\bd{\hat{\theta}}_k) =  \prod_{n: \,y_{sn}, \bd{x}_{sn} \mbox{\scriptsize{not missing}}}f_{nk}(y_{sn}|\bd{x}_{sn},\bd{\hat{\theta}}_{nk}).\] 

\subsection{Variable Selection Framework}\label{sec:varsel}
The variable selection procedure proposed in \citet{raftery06} and \citet{dean10} is ideal for application to the GMM. It reduces the variable selection problem to considering whether a single variable is useful for clustering or not. This is then combined with a search procedure (e.g. stepwise, backward, forward, headlong, etc.) to select a subset of the original variables as useful for clustering. The resulting framework allows for simultaneous variable selection and cluster estimation which is generally preferable to filter approaches. 

\subsubsection{Variable Selection Comparison Models}\label{sec:varselmod}
To check if a proposal variable is useful for clustering, two opposing models are posited (useful  for clustering versus not) and the fit compared to see which model has stronger evidence. For each stage we have three (potential) sets in a partition of the variables/measurements $\bd{y}$:
\begin{itemize}
\item $y^{(proposal)}$ - the (single) variable being proposed as a clustering variable
\item $\bd{y}^{(current)}$ - the current set of selected clustering variables 
\item $\bd{y}^{(not\, selected)}$ - all other variables (not being proposed or currently selected as the clustering variable).
\end{itemize}
The model for $y^{(proposal)}$ being useful for clustering, $M_1$, is a product of two sub-models:
\[  M_1(\bd{y}) = M_{clust}(y^{(proposal)},\bd{y}^{(current)})\times M_{not\,clust}(\bd{y}^{(not\, selected)}|y^{(proposal)},\bd{y}^{(current)}),\]
where $M_{clust}$ indicates a clustering model was fitted to the set of variables in parentheses and $M_{not\,clust}$ is a non-clustering model.

The model for $y^{(proposal)}$ \emph{not} being useful for clustering, $M_2$, is a product of three sub-models:
\[ M_2(\bd{y}) =  M_{clust}(\bd{y}^{(current)})\times M_{not\, clust}(y^{(proposal)}|\bd{y}^{(current)})\times M_{not\,clust}(\bd{y}^{(not\, selected)}|y^{(proposal)},\bd{y}^{(current)}).\]


Different approaches have been taken with respect to the sub-model, $M_{not\, clust}(y^{(proposal)}|\bd{y}^{(current)})$, for the relationship between the proposal variable and the current clustering variables in the model where the proposal variable does not have a clustering role. 
\begin{enumerate}
\item In the original paper, \citet{raftery06}, where model-based clustering with Gaussian components was considered, this took the form of a linear regression of $y^{(proposal)}$ on the full set of current clustering variables $\bd{y}^{(current)}$. 

\item In the \citet{dean10} paper, because conditional independence was assumed in the LCA model \citep{lazarsfeld68} applied to the categorical variables under consideration, it made sense to assume complete independence of $y^{(proposal)}$ from $\bd{y}^{(current)}$. This results in a reduction of  $M_{not\, clust}(y^{(proposal)}|\bd{y}^{(current)})$ to $M_{not\, clust}(y^{(proposal)})$. 

\item A compromise between these two extremes was proposed by \citet{maugis09} where variable selection was run on the regression of  $y^{(proposal)}$ on the full set of current clustering variables $\bd{y}^{(current)}$ (allowing $y^{(proposal)}$ to depend on all, a subset of or none of the set $\bd{y}^{(current)}$). \label{maugis-model}
\end{enumerate}
The modeling choices for the GMM case for $M_{not\, clust}(y^{(proposal)}|\bd{y}^{(current)})$, based on approach \ref{maugis-model}, are discussed further in Section \ref{sec:varselgmm}.

For a proposed variable $y^{(proposal)}$, once models $M_1$ and $M_2$ have been estimated from the data, a method for evaluating the evidence for one versus the other is needed. The obvious choice for doing so is to examine the Bayes factor of model 1 versus model 2, $B_{12}$
\[ B_{12} = \frac{p(\bd{y}|M_1)}{p(\bd{y}|M_2)}, \]
where $p(\bd{y}|M_i)$ is the integrated likelihood of model $M_i$. These integrated likelihoods are not available in closed form for finite mixture models. So instead, an approximation of the Bayes factor using the BIC is implemented.
\begin{equation} \log(B_{12}) \approx BIC(M_1) - BIC(M_2). \label{eqn:bicdiff}\end{equation}
As lower values of the BIC, as defined in equation \eqref{eqn:bic}, indicate a better fit,  if the difference in BIC values in equation \eqref{eqn:bicdiff} is negative, this indicates more evidence in support of model $M_1$ than $M_2$, suggesting the proposal variable, $y^{(proposal)}$ is useful for clustering. Conversely, if the difference is positive, there is more evidence supporting $M_2$ than $M_1$, suggesting the  proposal variable, $y^{(proposal)}$ is not useful for clustering. The natural default value of the threshold for the BIC difference for deciding if $y^{(proposal)}$ should be included in the set of clustering variables is 0. This value can be altered if stronger evidence is believed to be necessary before inclusion of a proposal variable into the set of clustering variables.

For $M_1$ we have:
\[ BIC(M_1) = BIC( M_{clust}(y^{(proposal)},\bd{y}^{(current)}))+ BIC(M_{not\,clust}(\bd{y}^{(not\, selected)}|y^{(proposal)},\bd{y}^{(current)})) \]
For $M_2$ we have:
\begin{eqnarray*}  BIC(M_2) &=& BIC(M_{clust}(\bd{y}^{(current)})) + BIC(M_{not\, clust}(y^{(proposal)}|\bd{y}^{(current)})) \\
&&+ BIC(M_{not\,clust}(\bd{y}^{(not\, selected)}|y^{(proposal)},\bd{y}^{(current)})) \end{eqnarray*}
We see that $M_{not\,clust}(\bd{y}^{not\, selected}|y^{(proposal)},\bd{y}^{(current)})$ appears in both models so this cancels in the difference giving us:
\begin{align}  
 BIC_{\mathit{diff}}(y^{(proposal)}) &= BIC(M_{clust}(y^{(proposal)},\bd{y}^{(current)}))
 \\ 
& - \left( BIC(M_{clust}(\bd{y}^{(current)})) + BIC(M_{not\, clust}(y^{(proposal)}|\bd{y}^{(current)}))\right) \nonumber  \end{align}

So, for each proposed, $y^{(proposal)}$, the cluster model is fit on the set of variables including the current set of clustering variables $\bd{y}^{(current)}$ along with the proposal variable $y^{(proposal)}$ and the BIC score for that model, $BIC(M_1)$, is calculated. The cluster model is fit on only the current set of clustering variables $\bd{y}^{(current)}$ and the BIC score for this model is calculated, $BIC(M_{clust}(\bd{y}^{(current)}))$, and either the regression (with or without variable selection) is fit for $y^{(proposal)}$ on $\bd{y}^{(current)}$ or a single component cluster model is fit on $y^{(proposal)}$ and $BIC(M_{not\, clust}(y^{(proposal)}|\bd{y}^{(current)}))$ or $BIC(M_{not\,clust}(\bd{y}^{(current)}))$ is produced. These are plugged into the $BIC_{\mathit{diff}}(y^{(proposal)})$ to give a value that can be used to make a decision about the proposed clustering variable. This is then combined with a search algorithm, such as those presented in the next section, to produce a set of selected clustering variables. 

\subsubsection{Variable Selection Search Algorithm}\label{sec:varselsearch}
Given the nature of the data examined where variables correspond to repeated measurements or time points, the number of variables is not expected to be large. This means that there will be no issues with fitting a GMM using the full set of measurements. Therefore we describe two backward search algorithms that will be used in producing the results for Section \ref{sec:results}. \\
\\
\noindent\textbf{\underline{Basic Greedy Backward Search}} \\
\\
The standard greedy backward search proceeds as follows:
\begin{enumerate}
\item Start with all variables in the $\bd{y}^{(current)}$ set
\item Take each variable from $\bd{y}^{(current)}$ individually in turn as $y^{(proposal)}$:
\begin{itemize}
\item Fit models $M_1$ and $M_2$
\item Calculate $BIC_{\mathit{diff}}$ using equation (8) 
\end{itemize}
\item Choose the variable with largest $BIC_{\mathit{diff}}$ value
\item If the variable's $BIC_{\mathit{diff}}$ is larger than the chosen threshold (usually 0), then remove this variable from the set of clustering variables and return to step 2. Otherwise, halt the algorithm.
\end{enumerate}
\noindent\textbf{\underline{Greedy Backward Search with Monotonicity}} \\
\\
Given that one of the common forms of data to which GMM is applied is repeated measurements where there is a temporal ordering to the recording of the repeated measurements, a greedy backward search with monotonicity may be of interest. 

This type of search proceeds as follows:
\begin{enumerate}
\item Start with all variables in the $\bd{y}^{(current)}$ set
\item Take the \emph{earliest} and the \emph{latest} variable from $\bd{y}^{(current)}$ individually in turn as $y^{(proposal)}$:
\begin{itemize}
\item Fit models $M_1$ and $M_2$
\item Calculate $BIC_{\mathit{diff}}$ using equation (8) 
\end{itemize}
\item Choose the variable with largest $BIC_{\mathit{diff}}$ value
\item If the variable's $BIC_{\mathit{diff}}$ is larger than the chosen threshold (usually 0), then remove this variable from the set of clustering variables and return to step 2. Otherwise, halt the algorithm.
\end{enumerate}

The backward greedy search in the case of limited number of variables/measurements should be the most efficient (where there is at least some clustering going on). However, it is perfectly feasible to look at forward or forward-and-backward stepwise search types as well as headlong (as described in \citet{dean10}) instead of greedy approaches as wrappers for the framework from Section \ref{sec:varselmod}.

\subsection{Growth Mixture Model with Variable Selection}\label{sec:varselgmm}
This section presents the details for model $M_1(\mathbf{y})$, where $y^{(proposal)}$ is useful for clustering and $M_2(\mathbf{y})$, where $y^{(proposal)}$ is {\it not} useful for clustering in the framework of growth mixture modeling with variable selection.   As discussed in Section \ref{sec:varselmod}, we can ignore $M_{not\,clust}(\bd{y}^{not\, selected}|y^{(proposal)},\bd{y}^{(current)})$ as it appears in both models.  We focus on $M_{clust}(y^{(proposal)},\bd{y}^{(current)})$, $M_{clust}(\bd{y}^{(current)})$ and $M_{not\, clust}(y^{(proposal)}|\bd{y}^{(current)})$.   Both $M_{clust}$ models come directly from Equations \eqref{eqn:gmm} and \eqref{eqn:gmmreg} depending on whether or not we have the inclusion of covariates.

Let $\mathcal{V}$ be the set of indices from $N_s$ for the current clustering variables.  For example, if there were 5 repeated measurements and the third and fifth are current clustering variables, then $\mathcal{V} = \{3, 5\}$.   Additionally, let $p$ represent the index of the proposed clustering variable.

Without covariates, we have 

$$M_{clust}(y^{(proposal)},\bd{y}^{(current)}) =\prod \limits_{s=1}^S \sum \limits_{k=1}^K \pi_k \prod \limits_{v  \in  \mathcal{V} \cup p} f_{vk}(y_{sv}|\bd{\theta}_{vk}),$$

and

$$M_{clust}(\bd{y}^{(current)}) = \prod \limits_{s=1}^S \sum \limits_{k=1}^K \pi_k \prod \limits_{v  \in  \mathcal{V}} f_{vk}(y_{sv}|\bd{\theta}_{vk}).$$

For both of these clustering models, $f_{vk}$ is Gaussian with $\bd{\theta}_{vk} = (\bd{\mu}_{vk}, \bd{\Sigma}_{vk})$.

The addition of covariates gives us

$$M_{clust}(y^{(proposal)},\bd{y}^{(current)}) = \prod \limits_{s=1}^S \sum \limits_{k=1}^K \pi_k \prod \limits_{v  \in  \mathcal{V} \cup p} f_{vk}(y_{sv}|\bd{x}_{sv}, \bd{\theta}_{vk}),$$

and

$$M_{clust}(\bd{y}^{(current)}) = \prod \limits_{s=1}^S \sum \limits_{k=1}^K \pi_k \prod \limits_{v  \in  \mathcal{V}} f_{vk}(y_{sv}|\bd{x}_{sv}, \bd{\theta}_{vk}),$$

where again, $f_{vk}$ is Gaussian, but for this finite mixture of regression models, $\bd{\theta}_{vk}$ is a set of component specific regression parameters as described in Section \ref{sec:fmm}.

The model for  $M_{not\, clust}(y^{(proposal)}|\bd{y}^{(current)})$ without covariates is given by

$$M_{not\, clust}(y^{(proposal)}|\bd{y}^{(current)}) = \prod \limits_{s=1}^S f_p(y_{s}^{(proposal)}|\bd{y}_s^{(current)*},\bd{\theta}_p), $$

where $\bd{y}^{(current)*}$ is a selection of $\bd{y}^{(current)}$ chosen using a regression variable selection based on BIC \citep{raftery95}.

The addition of covariates requires us to first fit the regression of $y^{(proposal)}$ on $\bd{x}^{(proposal)}$, we then select $\bd{y}^{(current)*}$ from the residuals of the previous regression on $\bd{y}^{(current)}$.  Giving a final model of 

$$M_{not\, clust}(y^{(proposal)}|\bd{y}^{(current)}) = \prod \limits_{s=1}^S f_p(y_{s}^{(proposal)}|\bd{x}_s^{(proposal)}, \bd{y}_s^{(current)*},\bd{\theta}_p).$$

\section{Results}\label{sec:results}
The starting values for EM estimation in all the following sections were produced using the k-means algorithm on all $\bd{y}$ values for the measurements being considered in the model. Using 50 random initializations, this was found to produce reasonable sets of starting values in general, and was computationally quick. Other starting value schemes could, of course, be incorporated.

Only complete data (no missing values) were generated in the simulations in Section \ref{sec:simstudy} and complete cases were used in the dataset in Section \ref{sec:data}. The presented methodology can be extended to include missing values using the ideas in Section \ref{sec:missdata}, but evaluating the impact of this was not considered to be the goal of this paper.

\subsection{Simulation Study}\label{sec:simstudy}

For each simulated data set, we compare the final GMM after variable selection with the GMM using all of the variables.  We can compare the accuracy of clustering  by looking at the difference in the Adjusted Rand Index (ARI, \citet{hubert85}) compared to the true simulated partition for each model.  Additionally, we can compare the accuracy of estimation by looking at the difference in the Root Mean Square Error (RMSE, \citet{steel60}) for each model.  We include the RMSE for the full set of variables based on the model using the selected variables and the model using all variables. Note that the RMSE for the model estimated using selected variables includes estimation of the non-selected clustering variables assuming a 1 group model.  We also provide results on the number of non-clustering variables chosen and the number of clusters chosen by each model.

All results are for 20 time points with 3 groups. 
 (Twenty time points was chosen as a reasonable upper limit for the number of variables that would be expected in a typical application. The method will, of course, run faster on a smaller number of time points.) The non-clustering time points all have intercept 0 and slope 1 with standard deviation 0.5 and standard normal distributed explanatory variables. All simulations are repeated 50 times and average results are reported in Tables \ref{tab1}, \ref{tab2}, \ref{tab3} and \ref{tab4}.

Tables \ref{tab1}, \ref{tab2} and \ref{tab3} all refer to clusterings with roughly equal sized clusters (in terms of membership/mixture probabilities) with varying degrees of separation in the two clustering time points.   In the first simulation summarized in Table \ref{tab1}, the first clustering variable has slope parameters (1, 3, -2) and the second clustering variable has slope parameters (1, 2.5, -0.5)  over the three groups respectively.    98\% of the simulations correctly selected both the clustering variables and 43\% of the simulations correctly chose 3 groups.  Additionally, 78\% of the simulations had higher ARI and 66\% of the simulations had lower RMSE for the model using only the selected clustering variables.  In the second and third simulation, both clustering variables share the same slopes for the three groups, with Table \ref{tab2} having slopes (1, 2.5, -0.5)  and Table \ref{tab3} having slopes (1, 3, -2).    We see an improvement in all of our measures of evaluation for these two simulations.  

Finally in Table \ref{tab4}, the membership/mixture probabilities place 70\% of the observations in one group and split the remaining 30\% into the other two groups.  The slopes for the clustering variables match that of the third simulation.  Even with unbalanced groups, the clustering variables are selected correctly 96\% of the time and three groups are chosen 56\% of the time.  It is worth noting that clustering of all the time points resulted in only 2 groups being chosen for every repetition of the simulation, demonstrating the detrimental effect that the inclusion of non-clustering variables can have on cluster estimation.  The ARI was higher for 72\% of the simulations and the RMSE was smaller for 90\% of the simulations for the model using only the selected clustering variables.

Overall, as can be seen from the tables, the variable selection procedure almost always selects the clustering variables, but also often includes some additional variables. However, as can be seen from the ARI, in terms of group structure recovery, most of the time it is better to just select the cluster model on the basis of the selected variables rather than the full variable set. Similarly, if good estimation performance is the goal, again in terms of RMSE, it is usually better to fit the model based on the selected variables alone. 
\begin{table}[htbp]
\begin{tabular}{lrrr}
\hline
\hline
\multicolumn{4}{c}{\textbf{Simulation parameters}} \\
Group mixture weights&Group 1 = 0.3& Group 2 = 0.3& Group 3 = 0.4 \\
Clustering time points&5 and 15&& \\
\hline
Clustering time points slopes&Group 1&Group 2&Group 3 \\
\hline
Time point 5&1&3&-2 \\
Time point 15&1&2.5&-0.5 \\
\hline
\hline
\multicolumn{4}{c}{\textbf{Simulation results}} \\
Clustering time point selected&Time point 5& Time point 15&Both time points \\
\hline
\% simulations&98\%&100\%&98\% \\
\hline
\end{tabular}
\begin{tabular}{lrrrrrrrrr}
\hline
Number of non-clustering&&&&&&&&& \\
time points selected&0&1&2&3&4&6&7&8&17 \\
\hline
\% simulations&26\%&26\%&6\%&8\%&8\%&8\%&12\%&4\%&2\% \\
\hline
\hline
Number of groups chosen&2&3&4&&&&&& \\
\hline
\% of simulations for &&&&&&&&& \\
model with \emph{selected} variables&14\%&42\%&44\%&&&&&& \\ 
\% of simulations for&&&&&&&&& \\
model with \emph{all} variables&70\%&12\%&18\%&&&&&& \\
\hline
\end{tabular}
\begin{tabular}{rrrrrrr}
\hline
\multicolumn{7}{c}{Summary statistics for difference between ARI for clustering with selected  } \\
\multicolumn{7}{c}{variables versus clustering with all variables} \\
Minimum&1$^{st}$ quartile&Median&Mean&3$^{rd}$ quartile&Maximum& \\
\hline
-0.4969&	0.01145	&0.15330&	0.18270&	0.30260&0.7489& \\
\hline
\multicolumn{7}{l}{78\% of simulations had a higher ARI for the model using only the } \\
\multicolumn{7}{l}{selected variables} \\
\hline
\hline
\multicolumn{7}{c}{Summary statistics for difference between RMSE for clustering with selected} \\
\multicolumn{7}{c}{ variables versus clustering with all variables} \\
Minimum&1$^{st}$ quartile&Median&Mean&3$^{rd}$ quartile&Maximum& \\
\hline
-0.64120    &	-0.22790	&-0.11790	&-0.12330	&0.04693&0.92630& \\
\hline
\multicolumn{7}{l}{66\% of simulations had lower RMSE for all time points for the model using only the } \\
\multicolumn{7}{l}{selected variables} \\
\hline
\end{tabular}
\caption{First simulations set}
\label{tab1}
\end{table}

\begin{table}[htbp]
\begin{tabular}{lrrr}
\hline
\hline
\multicolumn{4}{c}{\textbf{Simulation parameters}} \\
Group mixture weights&Group 1 = 0.3& Group 2 = 0.3& Group 3 = 0.4 \\
Clustering time points&5 and 15&& \\
\hline
Clustering time points slopes&Group 1&Group 2&Group 3 \\
\hline
Time point 5&1&2.5&-0.5 \\
Time point 15&1&2.5&-0.5 \\
\hline
\hline
\multicolumn{4}{c}{\textbf{Simulation results}} \\
Clustering time point selected&Time point 5& Time point 15&Both time points \\
\hline
\% simulations&100\%&100\%&100\% \\
\hline
\end{tabular}
\begin{tabular}{lrrrrrrrrr}
\hline
Number of non-clustering&&&&&&&&& \\
time points selected&0&1&2&3&4&5&6&14&17 \\
\hline
\% simulations&66\%&20\%&6\%&2\%&1\%&1\%&2\%&1\%&1\% \\
\hline
\hline
Number of groups chosen&2&3&4&&&&&& \\
\hline
\% of simulations for &&&&&&&&& \\
model with \emph{selected} variables&19\%&52\%&29\%&&&&&& \\ 
\% of simulations for&&&&&&&&& \\
model with \emph{all} variables&94\%&6\%&0\%&&&&&& \\
\hline
\end{tabular}
\begin{tabular}{rrrrrrr}
\hline
\multicolumn{7}{c}{Summary statistics for difference between ARI for clustering with selected  } \\
\multicolumn{7}{c}{variables versus clustering with all variables} \\
Minimum&1$^{st}$ quartile&Median&Mean&3$^{rd}$ quartile&Maximum& \\
\hline
-0.428100	&-0.002086&	0.063560	&0.050580	&0.130200	&0.275100 \\
\hline
\multicolumn{7}{l}{77\% of simulations had a higher ARI for the model using only the } \\
\multicolumn{7}{l}{selected variables} \\
\hline
\hline
\multicolumn{7}{c}{Summary statistics for difference between RMSE for clustering with selected} \\
\multicolumn{7}{c}{ variables versus clustering with all variables} \\
Minimum&1$^{st}$ quartile&Median&Mean&3$^{rd}$ quartile&Maximum& \\
\hline
-0.5284&	-0.3689	&-0.3232&	-0.2634	&-0.1757	&0.3829 \\
\hline
\multicolumn{7}{l}{94\% of simulations had lower RMSE for all time points for the model using only the } \\
\multicolumn{7}{l}{selected variables} \\
\hline
\end{tabular}
\caption{Second simulations set}
\label{tab2}
\end{table}

\begin{table}[htbp]
\begin{tabular}{lrrr}
\hline
\hline
\multicolumn{4}{c}{\textbf{Simulation parameters}} \\
Group mixture weights&Group 1 = 0.3& Group 2 = 0.3& Group 3 = 0.4 \\
Clustering time points&5 and 15&& \\
\hline
Clustering time points slopes&Group 1&Group 2&Group 3 \\
\hline
Time point 5&1&3&-2 \\
Time point 15&1&3&-2 \\
\hline
\hline
\multicolumn{4}{c}{\textbf{Simulation results}} \\
Clustering time point selected&Time point 5& Time point 15&Both time points \\
\hline
\% simulations&100\%&100\%&100\% \\
\hline
\end{tabular}
\begin{tabular}{lrrrrrrrrr}
\hline
Number of non-clustering&&&&&&&&& \\
time points selected&0&1&2&3&4&5&6&7& \\
\hline
\% simulations&45\%&23\%&7\%&7\%&7\%&2\%&5\%&5\%& \\
\hline
\hline
Number of groups chosen&2&3&4&&&&&& \\
\hline
\% of simulations for &&&&&&&&& \\
model with \emph{selected} variables&2\%&77\%&20\%&&&&&& \\ 
\% of simulations for&&&&&&&&& \\
model with \emph{all} variables&61\%&34\%&5\%&&&&&& \\
\hline
\end{tabular}
\begin{tabular}{rrrrrrr}
\hline
\multicolumn{7}{c}{Summary statistics for difference between ARI for clustering with selected  } \\
\multicolumn{7}{c}{variables versus clustering with all variables} \\
Minimum&1$^{st}$ quartile&Median&Mean&3$^{rd}$ quartile&Maximum& \\
\hline
-0.15790&	-0.01606	&0.17880&	0.13660&	0.21700&	0.55370& \\
\hline
\multicolumn{7}{l}{73\% of simulations had a higher ARI for the model using only the } \\
\multicolumn{7}{l}{selected variables} \\
\hline
\hline
\multicolumn{7}{c}{Summary statistics for difference between RMSE for clustering with selected} \\
\multicolumn{7}{c}{ variables versus clustering with all variables} \\
Minimum&1$^{st}$ quartile&Median&Mean&3$^{rd}$ quartile&Maximum& \\
\hline
-1.143000	&-0.531900&	-0.479300	&-0.388600&	-0.047100&	-0.005398& \\
\hline
\multicolumn{7}{l}{100\% of simulations had lower RMSE for all time points for the model using only the } \\
\multicolumn{7}{l}{selected variables} \\
\hline
\end{tabular}
\caption{Third simulations set}
\label{tab3}
\end{table}

\begin{table}[htbp]
\begin{tabular}{lrrr}
\hline
\hline
\multicolumn{4}{c}{\textbf{Simulation parameters}} \\
Group mixture weights&Group 1 = 0.7& Group 2 = 0.15& Group 3 = 0.15 \\
Clustering time points&5 and 15&& \\
\hline
Clustering time points slopes&Group 1&Group 2&Group 3 \\
\hline
Time point 5&1&2.5&-0.5 \\
Time point 15&1&2.5&-0.5 \\
\hline
\hline
\multicolumn{4}{c}{\textbf{Simulation results}} \\
Clustering time point selected&Time point 5& Time point 15&Both time points \\
\hline
\% simulations&96\%&96\%&96\% \\
\hline
\end{tabular}
\begin{tabular}{lrrrrrrrrr}
\hline
Number of non-clustering&&&&&&&&& \\
time points selected&0&1&2&3&4&6&17& \\
\hline
\% simulations&54\%&22\%&10\%&6\%&4\%&2\%&2\%&& \\
\hline
\hline
Number of groups chosen&1&2&3&4&&&&& \\
\hline
\% of simulations for &&&&&&&&& \\
model with \emph{selected} variables&4\%&20\%&56\%&20\%&&&&& \\ 
\% of simulations for&&&&&&&&& \\
model with \emph{all} variables&0\%&100\%&0\%&0\%&&&&& \\
\hline
\end{tabular}
\begin{tabular}{rrrrrrr}
\hline
\multicolumn{7}{c}{Summary statistics for difference between ARI for clustering with selected  } \\
\multicolumn{7}{c}{variables versus clustering with all variables} \\
Minimum&1$^{st}$ quartile&Median&Mean&3$^{rd}$ quartile&Maximum& \\
\hline
-0.199900&	-0.001656	&0.106100&	0.117800&	0.211500&	0.609400 & \\
\hline
\multicolumn{7}{l}{72\% of simulations had a higher ARI for the model using only the } \\
\multicolumn{7}{l}{selected variables} \\
\hline
\hline
\multicolumn{7}{c}{Summary statistics for difference between RMSE for clustering with selected} \\
\multicolumn{7}{c}{ variables versus clustering with all variables} \\
Minimum&1$^{st}$ quartile&Median&Mean&3$^{rd}$ quartile&Maximum& \\
\hline
-0.27330	&-0.15030&	-0.10200	&-0.09932&	-0.05620&	0.08910& \\
\hline
\multicolumn{7}{l}{90\% of simulations had lower RMSE for all time points for the model using only the } \\
\multicolumn{7}{l}{selected variables} \\
\hline
\end{tabular}
\caption{Fourth simulations set}
\label{tab4}
\end{table}

\subsection{Application}\label{sec:data}

In this section, we apply variable selection to the Pittsburgh 600, a  longitudinal data set composed of five depression studies, each with a different treatment protocol (\citet{thase1997}).  Subjects were diagnosed as clinically depressed upon entrance to the study and there are 26 weekly scores indicating their current level of depression.  In addition to their score, we have several explanatory variables, including the subjects' sex, age and medication status.

\begin{figure}[htbp]
\begin{center}
\includegraphics[scale=0.5]{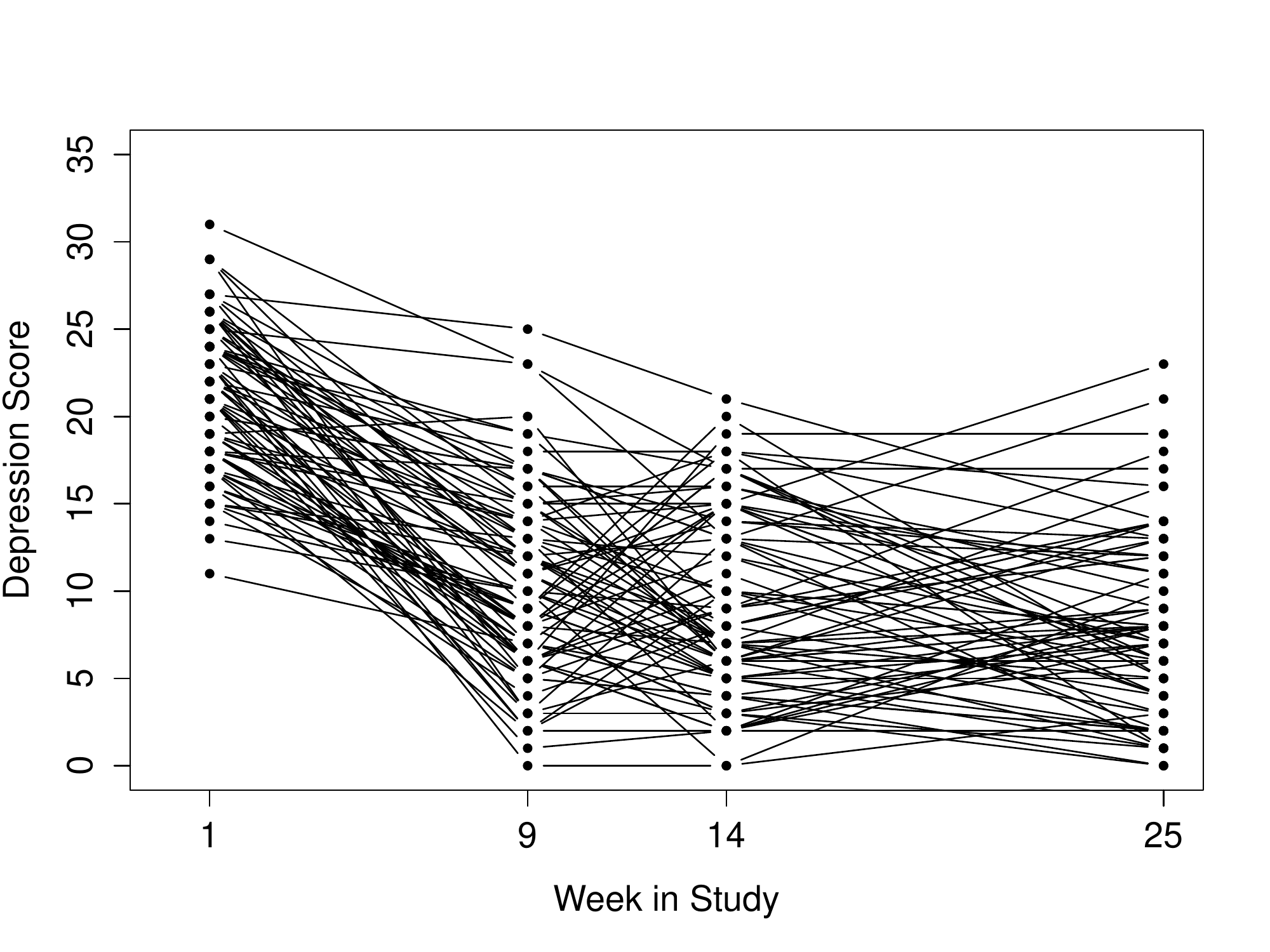}
\end{center}
\caption{Selected subjects in Pittsburgh 600 study}\label{fig:trajectories}
\end{figure}

Because most trajectories in this study have intermittent missingness (which was not incorporated into our method, as it was not the focus of this manuscript), we have taken a subset of four time points that provided a large number of complete trajectories.  There were 74 subjects that had complete observations at Weeks 1, 9, 14 and 25 of the study.   Figure~\ref{fig:trajectories} shows the trajectories of these subjects over time.  We can see that overall there is a decreasing trend, showing a reduction in clinical depression scores.

\begin{figure}[htbp]
\begin{center}
\includegraphics[scale=0.5]{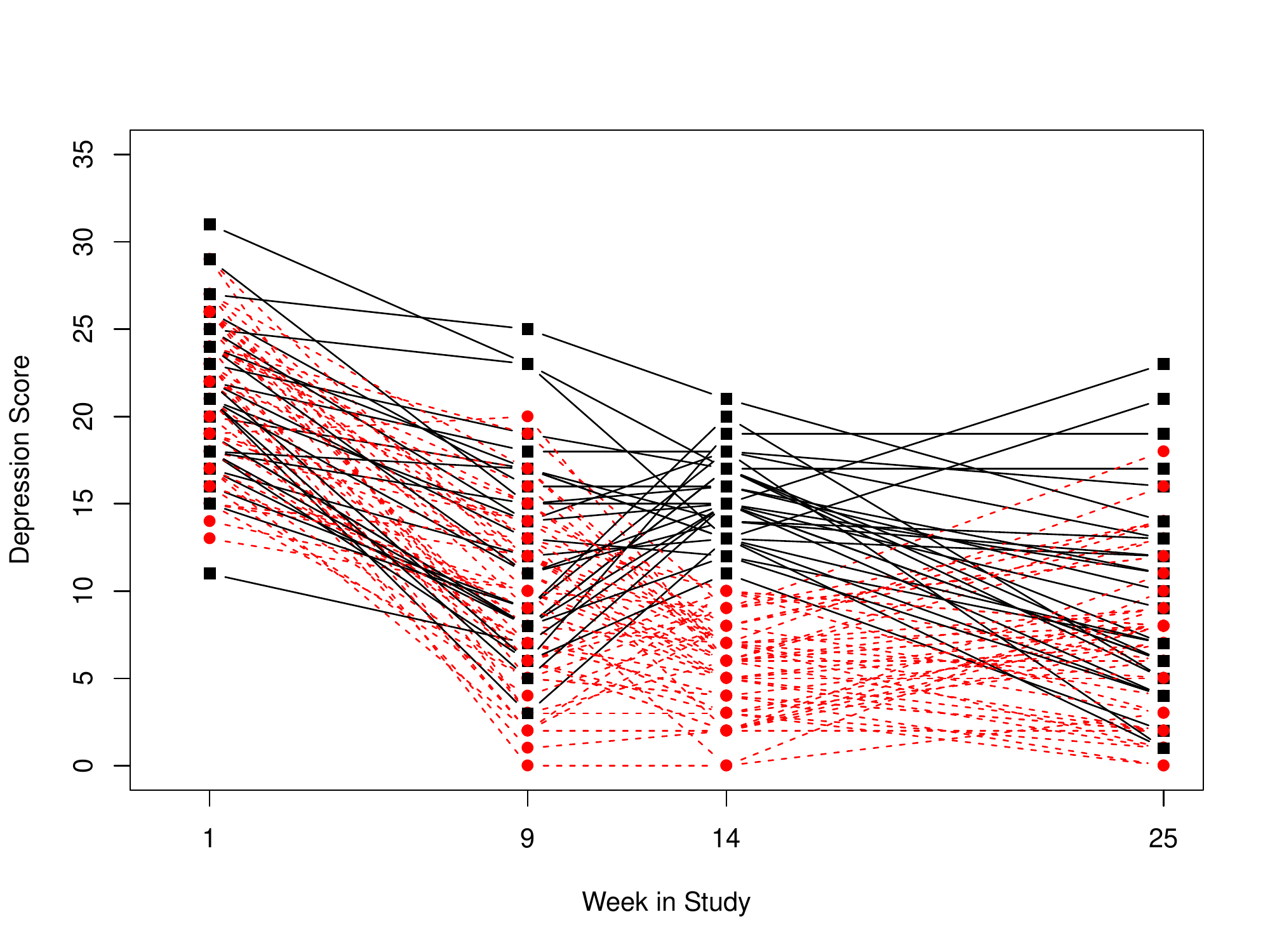}
\end{center}
\caption{Clusters resulting from GMM fit only to Week 14}\label{fig:clusters}
\end{figure}

\begin{figure}[htbp]
\begin{center}
\includegraphics[scale=0.5]{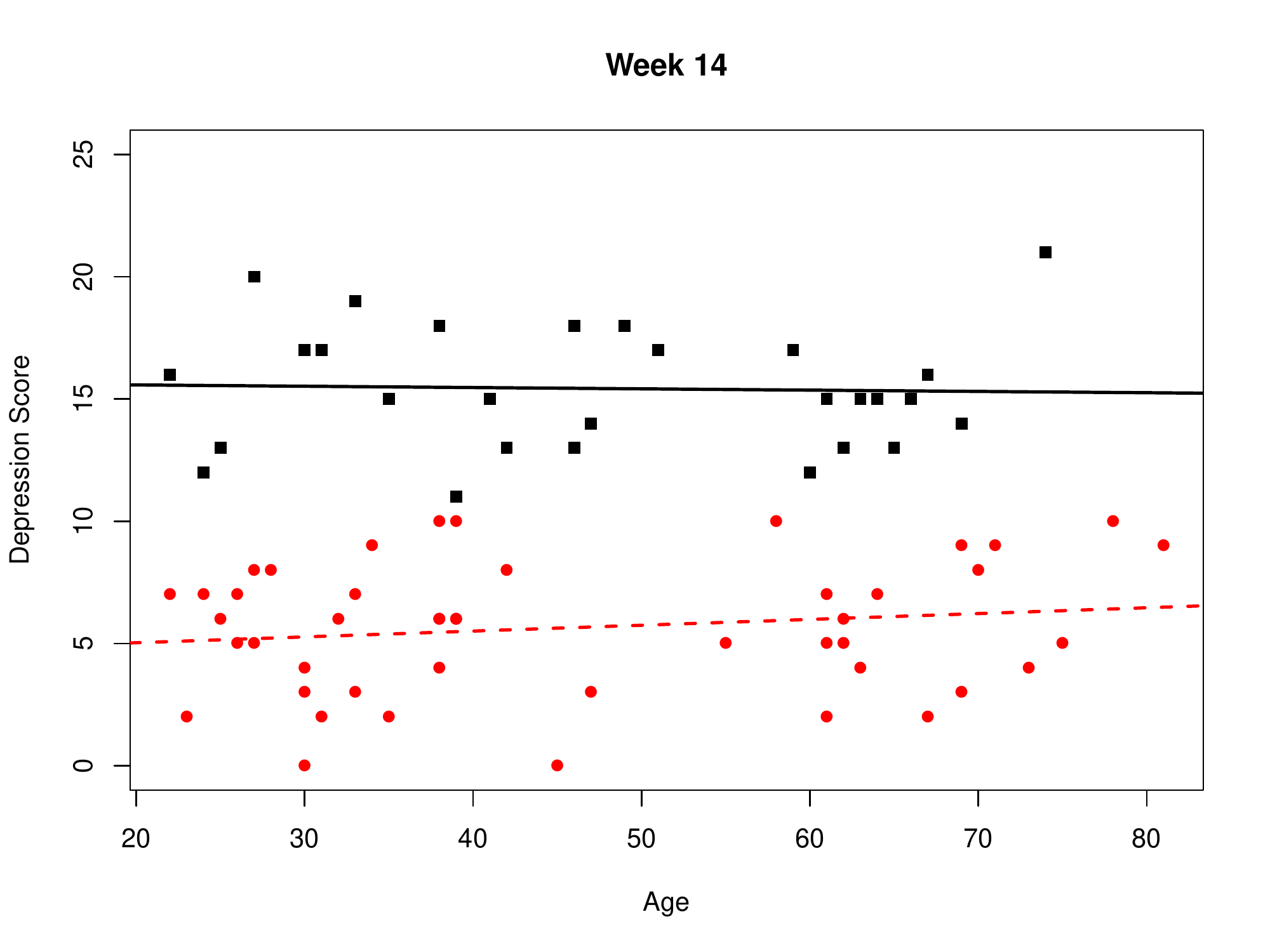}
\end{center}
\caption{Clusters at Week 14}\label{fig:week14}
\end{figure}

Using subject age as the explanatory variable, variable selection chose Week 14 as the only time point that is useful for clustering.  Clustering on Week 14 produces two classes, containing 46 and 28 trajectories as seen in Figure~\ref{fig:clusters}.  In Figure~\ref{fig:week14}, we see that the two classes obtained from the growth mixture model are really driven by different intercepts.

\begin{figure}[htbp]
\begin{center}
\includegraphics[scale=0.5]{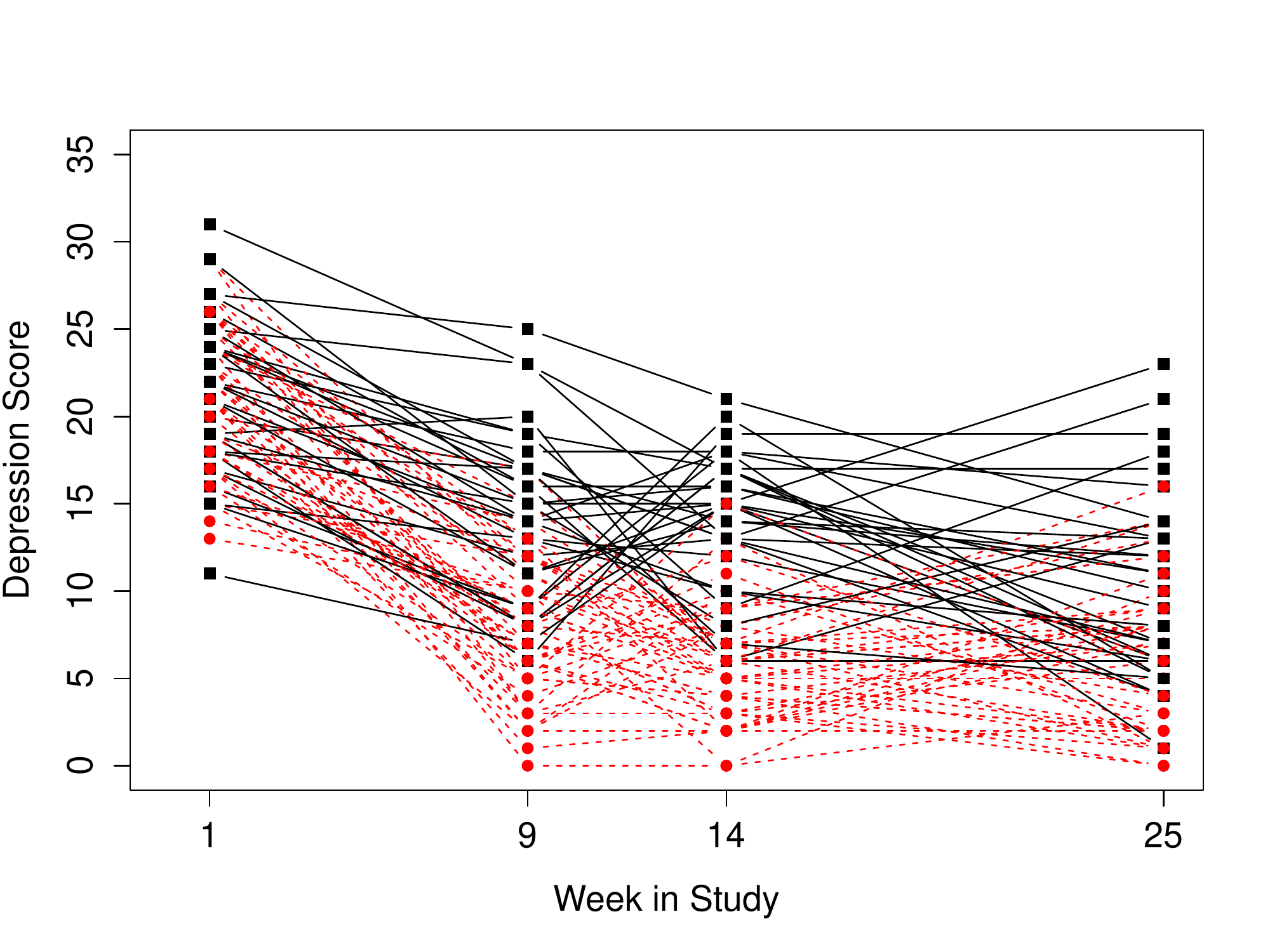}
\end{center}
\caption{Clusters resulting from GMM fit to all four time points}\label{fig:fullclustering}
\end{figure}

When a growth mixture model is fit to all 4 time points, 2 groups of size 43 and 31 are selected.  The clustering can be seen in Figure~\ref{fig:fullclustering}.   When we compare these two clustering solutions, the ARI is 0.487 indicating that the solutions are not very similar. It is worth noting that we are seeing nearly identical results from the variable selection, growth mixture model procedure when the explanatory variable is changed to whether or not each subject is on a medication and also which of the 5 studies they belong to.  Using all of the time points for the growth mixture model with either alternative explanatory variable however, produces more variation between the clusterings.

\section{Discussion}\label{sec:discuss}
This paper presents a framework to incorporate variable selection into the growth mixture model for the settings where one is clustering repeated measurements over time, with possible covariates. The simulated results suggest from both an estimation and interpretation point of view, that it is preferable to apply variable selection when using a GMM. It results in better estimation of the number of groups present in the data and superior estimation of the parameters in the resulting regressions.

 A run on simulated data with 20 time points (2 representing true clustering variables), 3 groups and 400 datapoints takes 4987 seconds, i.e. just over an hour on an iMac with a 3.06 GHz Intel Core Duo with 8 GB 1067 MHz DDR3 memory. This represents software written only in R which has not been optimized for speed. Some of the routines  could be transferred to C++ and run much faster as a result. Since the search algorithm loops over the number of time points, datasets with smaller numbers of time points would be expected to take less time as well.

The clustering found using the selected variables can often vary from that found using the full original set of variables/measurements as seen in the application in Section \ref{sec:data}. Which cluster model is of more interest can depend on the context of the problem. If one is interested in the grouping on all variables, a priori, then the variable selection is interesting only in the sense of highlighting variables that best separate the groups, not in the sense of estimating the clustering itself. If this is so, then variable selection can be performed to find these variables but the clustering model estimation is best done on the full set. Or indeed an alternative post-hoc variable selection method could be used after the clustering has already been performed. In other situations, it may be the case that both variable selection and clustering on the selected variables is the goal. It may also still be of interest to compare the clustering found on the selected variables versus the clustering on the original full set of variables.

An added advantage of variable selection in the context of GMMs is that, given many of the associated degeneracy and local optima issues are related to dimensionality, any reduction in dimensions will hopefully see a reduction in the occurrence of such. Although, of course, it cannot guarantee that they will not occur.

The framework presented here is extremely flexible and can be extended to facilitate estimation using incomplete observations as mentioned previously. 

This framework could also be adapted to the generalized linear  model framework in addition to the linear model approach presented in this paper, or the generalized additive model clustering case as seen in \citet{james03}. It would also be possible to extend the model to allow an autoregressive time dependence instead of the conditional independence presented, either by using previous time points as covariates in the covariate GMM setting or using the framework presented in \citet{mcnicholas12}.

In addition to variable selection in the clustering sense, it would also be possible to incorporate variable selection in the regression for the outcome variables on their covariates in the GMM setting.

A classification or supervised version of our methodology would be easy to implement since it essentially involves an observed version of the missing data (group labels) in the EM, and a single M step for parameter estimation followed by a single E step for estimating group membership probabilities for new data. A hybrid semi-supervised approach, similar to \citet{murphy10}, could also be considered.

Recent work by \citet{baudry10}, \citet{hennig10}, \citet{scrucca16}, and \citet{Melnykov2016} have examined different ways of combining components to create multi-component clusters as opposed to the ``one component equals one cluster'' approach mentioned in section \ref{sec:mixclust}. This is useful in cases where the underlying cluster distribution is different from the assumed component distribution. Particularly in cases with skewness or heavy tails, setting each component as a cluster can lead to an overestimate of the number of groups in the data. It may be possible for this to happen in the GMM case as well, particularly when there are outliers, so future work could look at adapting the methods for combining marginal density components to combining conditional density components instead. 

Issues with this method can arise with difficulties in finding good starting values for the clustering. K-means does a good job in most cases but other methods could be used in place of this should problems arise. There can also be issues when a cluster is assigned to only one or a pair of trajectories as this can result in singular matrix issues during estimation of the cluster and regression parameters. Including a noise component in the clustering might be a possible method to deal with this issue (as a single observation cluster could be argued to be an outlier/noise observation). Bayesian estimation used as an alternative to the EM based estimation advocated in this paper, could also help with regularization issues.

This paper presents an initial look at variable selection in the growth mixture model framework, which will hopefully stimulate interesting further research in this area. R code for fitting the models described in this paper is available on request from the authors.

\bibliography{refs.bib}

\begin{thebibliography}{38}
\newcommand{\enquote}[1]{``#1''}
\expandafter\ifx\csname natexlab\endcsname\relax\def\natexlab#1{#1}\fi

\bibitem[{Baudry et~al.(2010)Baudry, Raftery, Celeux, Lo, and
  Gottardo}]{baudry10}
Baudry, J., Raftery, A.~E., Celeux, G., Lo, K., and Gottardo, R. (2010),
  \enquote{Combining mixture components for clustering,} \textit{Journal of
  Computational and Graphical Statistics}, 332--353.

\bibitem[{Biernacki and Govaert(1997)}]{biernacki97}
Biernacki, C. and Govaert, G. (1997), \enquote{Using the classification
  likelihood to choose the number of clusters,} \textit{Computing Science and
  Statistics}, 29, 451--457.

\bibitem[{Biernacki and Govaert(1999)}]{biernacki99}
--- (1999), \enquote{Choosing models in model-based clustering and discriminant
  analysis,} \textit{Journal of Statistical Computation and Simulation}, 64.

\bibitem[{Dean and Raftery(2010)}]{dean10}
Dean, N. and Raftery, A.~E. (2010), \enquote{Latent class analysis variable
  selection,} \textit{Annals of the Institute of Statistical Mathematics}, 62,
  11--35.

\bibitem[{Dempster et~al.(1977)Dempster, Laird, and Rubin}]{dempster77}
Dempster, A.~P., Laird, N.~M., and Rubin, D.~B. (1977), \enquote{Maximum
  likelihood from incomplete data via the EM algorithm,} \textit{Journal of the
  Royal Statistical Society. Series B (methodological)}, 1--38.

\bibitem[{Everitt et~al.(2011)Everitt, Landau, Leese, and Stahl}]{everitt11}
Everitt, B., Landau, S., Leese, M., and Stahl, D. (2011), \textit{Cluster
  analysis}, Wiley Series in Probability and Statistics, Chichester, UK: Wiley.

\bibitem[{Fraley and Raftery(1998)}]{fraley98}
Fraley, C. and Raftery, A.~E. (1998), \enquote{How many clusters? Which
  clustering method? Answers via model-based cluster analysis,} \textit{The
  Computer Journal}, 41, 578--588.

\bibitem[{Fraley and Raftery(2002)}]{fraley02}
--- (2002), \enquote{Model-based clustering, discriminant analysis, and density
  estimation,} \textit{Journal of the American Statistical Association},
  611--631.

\bibitem[{Fraley et~al.(2012)Fraley, Raftery, Murphy, and Scrucca}]{mclust}
Fraley, C., Raftery, A.~E., Murphy, T.~B., and Scrucca, L. (2012),
  \enquote{mclust Version 4 for R: normal mixture modeling for model-based
  clustering, classification, and density estimation,} Tech. Rep. 597,
  Department of Statistics, University of Washington.

\bibitem[{Gr{\"u}n and Leisch(2008)}]{grun08}
Gr{\"u}n, B. and Leisch, F. (2008), \textit{Recent advances in linear models
  and related areas: essays in honour of Helge Toutenburg}, Physica-Verlag HD,
  chap. Finite mixtures of generalized linear regression models.

\bibitem[{Gupta and Chen(2011)}]{gupta11}
Gupta, M.~R. and Chen, Y. (2011), \enquote{Theory and Use of the EM Algorithm,}
  \textit{Foundations and Trends{\textregistered} in Signal Processing}, 4,
  223--296.

\bibitem[{Hartigan(1975)}]{hartigan75}
Hartigan, J.~A. (1975), \textit{Clustering algorithms}, Wiley.

\bibitem[{Hartigan(1981)}]{hartigan81}
--- (1981), \enquote{Consistency of single linkage for high-density clusters,}
  \textit{Journal of the American Statistical Association}, 76, 388--394.

\bibitem[{Hennig(2010)}]{hennig10}
Hennig, C. (2010), \enquote{Methods for merging {G}aussian mixture components,}
  \textit{Advances in Data Analysis and Classification}, 3--34.

\bibitem[{Hubert and Arabie(1985)}]{hubert85}
Hubert, L. and Arabie, P. (1985), \enquote{Comparing partitions,}
  \textit{Journal of Classification}, 2, 193--218.

\bibitem[{James and Sugar(2003)}]{james03}
James, G.~M. and Sugar, C.~A. (2003), \enquote{Clustering for sparsely sampled
  functional data,} \textit{Journal of the American Statistical Association},
  98, 565--576.

\bibitem[{Keribin(2000)}]{keribin00}
Keribin, C. (2000), \enquote{Consistent estimation of the order of mixture
  models,} \textit{Sankhy\-{a}}, 62, 49--66.

\bibitem[{Lazarsfeld and Henry(1968)}]{lazarsfeld68}
Lazarsfeld, P.~F. and Henry, N.~W. (1968), \textit{Latent structure analysis},
  Houghton Mifflin.

\bibitem[{MacQueen(1967)}]{macqueen67}
MacQueen, J.~B. (1967), \enquote{Some methods for classification and analysis
  of multivariate observations,} in \textit{Proceedings of 5th Berkeley
  Symposium on Mathematical Statistics and Probability}, University of
  California Press.

\bibitem[{Maugis et~al.(2009)Maugis, Celeux, and Martin-Magniette}]{maugis09}
Maugis, C., Celeux, G., and Martin-Magniette, M.-L. (2009), \enquote{Variable
  selection for clustering with Gaussian mixture models,} \textit{Biometrics},
  65, 701--709.

\bibitem[{McLachlan and Krishnan(2008)}]{mclachlan08}
McLachlan, G.~J. and Krishnan, T. (2008), \textit{The {E}{M} algorithm and
  extensions}, Wiley.

\bibitem[{McNicholas and Subedi(2012)}]{mcnicholas12}
McNicholas, P. and Subedi, S. (2012), \enquote{Clustering gene expression time
  course data using mixtures of multivariate t-distributions,} \textit{Journal
  of Statistical Planning and Inference}, 5.

\bibitem[{Melnykov(2016)}]{Melnykov2016}
Melnykov, V. (2016), \enquote{Merging mixture components for clustering through
  pairwise overlap,} \textit{Journal of Computational and Graphical
  Statistics}, 24, 66--90.

\bibitem[{Murphy et~al.(2010)Murphy, Dean, and Raftery}]{murphy10}
Murphy, T.~B., Dean, N., and Raftery, A.~E. (2010), \enquote{Variable selection
  and updating in model-based discriminant analysis for high dimensional data
  with food authenticity applications,} \textit{Annals of Applied Statistics},
  4, 396--421.

\bibitem[{Muth{\'e}n and Shedden(1999)}]{muthen99}
Muth{\'e}n, B. and Shedden, K. (1999), \enquote{Finite mixture modeling with
  mixture outcomes using the EM algorithm,} \textit{Biometrics}, 55, 463--469.

\bibitem[{Pearson(1894)}]{pearson94}
Pearson, K. (1894), \enquote{Contribution to the mathematical theory of
  evolution,} \textit{Philosophical Transactions of the Royal Society of
  London, Series A}, 71.

\bibitem[{{R Core Team}(2015)}]{Rlanguage}
{R Core Team} (2015), \textit{R: A language and environment for statistical
  computing}, R Foundation for Statistical Computing, Vienna, Austria.

\bibitem[{Raftery(1995)}]{raftery95}
Raftery, A.~E. (1995), \enquote{Bayesian model selection in social research
  (with Discussion),} \textit{Sociological Methodology}, 111--196.

\bibitem[{Raftery and Dean(2006)}]{raftery06}
Raftery, A.~E. and Dean, N. (2006), \enquote{Variable selection for model-based
  clustering,} \textit{Journal of the American Statistical Association}, 101,
  168--178.

\bibitem[{Ram and Grimm(2009)}]{ram2009}
Ram, N. and Grimm, K.~J. (2009), \enquote{Methods and measures: Growth mixture
  modeling: A method for identifying differences in longitudinal change among
  unobserved groups,} \textit{International Journal of Behavioral Development},
  33, 565--576.

\bibitem[{Rusakov and Geiger(2005)}]{rusakov05}
Rusakov, D. and Geiger, D. (2005), \enquote{Asymptotic model selection for
  naive bayesian networks,} \textit{Journal of Machine Learning Research}, 6,
  1--35.

\bibitem[{Schwarz(1978)}]{schwarz78}
Schwarz, G.~E. (1978), \enquote{Estimating the dimension of a model,}
  \textit{Annals of Statistics}, 6, 461--464.

\bibitem[{Scrucca(2016)}]{scrucca16}
Scrucca, L. (2016), \enquote{Identifying connected components in {G}aussian
  finite mixture models for clustering,} \textit{Computational Statistics \&
  Data Analysis}, 93, 5--17.

\bibitem[{Steel and Torrie(1960)}]{steel60}
Steel, R. G.~D. and Torrie, J. (1960), \textit{Principles and Procedures of
  Statistics with Special Reference to the Biological Sciences}, McGraw Hill.

\bibitem[{Thase et~al.(1997)Thase, Greenhouse, Frank, Reynolds, Pilkonis,
  Hurley, Grochocinski, and Kupfer}]{thase1997}
Thase, M.~E., Greenhouse, J.~B., Frank, E., Reynolds, C.~F., Pilkonis, P.~A.,
  Hurley, K., Grochocinski, V., and Kupfer, D.~J. (1997), \enquote{Treatment of
  major depression with psychotherapy or psychotherapy-pharmacotherapy
  combinations,} \textit{Archives of General Psychiatry}, 54, 1009--1015.

\bibitem[{Titterington et~al.(1985)Titterington, Smith, Makov,
  et~al.}]{titterington85}
Titterington, D.~M., Smith, A.~F., Makov, U.~E., et~al. (1985),
  \textit{Statistical analysis of finite mixture distributions}, vol.~7, Wiley
  New York.

\bibitem[{Ward(1963)}]{ward63}
Ward, J.~H. (1963), \enquote{Hierarchical grouping to optimize an objective
  function,} \textit{Journal of the American Statistical Association}, 58,
  236--244.

\bibitem[{Wishart(1969)}]{wishart69}
Wishart, D. (1969), \enquote{Mode analysis: A generalization of nearest
  neighbor which reduces chaining effects,} in \textit{Numerical Taxonomy}, ed.
  Cole, A.~J., Academic Press, pp. 282--311.

\end{thebibliography}
\bibliographystyle{asa}

\end{spacing}
\end{document}